\documentclass[10pt,A4,conference]{IEEEtran}

\usepackage{geometry}
 \usepackage{authblk}           
\usepackage[T1]{fontenc}
\usepackage{cite}
\usepackage{balance}

\ifCLASSOPTIONcompsoc
 \usepackage[caption=false,font=normalsize,labelfont=sf,textfont=sf]{subfig}
\else
 \usepackage[caption=false,font=footnotesize]{subfig}
\fi

\usepackage{algorithm}
\usepackage[noend]{algpseudocode}                       
\usepackage{amssymb}
\usepackage{amsthm}
\usepackage{array}
\usepackage{graphicx}
\usepackage{lettrine}
\usepackage{epsf} 
\usepackage{psfrag}
\usepackage[usenames,dvipsnames]{pstricks}
\usepackage{pst-grad} 
\usepackage{pst-plot} 
\usepackage[space]{grffile} 
\usepackage{etoolbox} 
\usepackage{url} 

\usepackage{amsmath}

\begin{document}

\title{A Duty-Cycle-Efficient Synchronization \\Protocol for Slotted-Aloha in LoRaWAN}

\author[]{Amavi Dossa}
\author[]{El Mehdi Amhoud}
\affil[]{College of Computing, Mohammed VI Polytechnic University (UM6P), Benguerir, Morocco, \authorcr
emails: {\{amavi.dossa,elmehdi.amhoud\}@um6p.ma}}

\maketitle

\begin{abstract}
In the current context of massive IoT, the Pure-Aloha scheme used in LoRaWAN is reaching its limit, and Slotted-Aloha is being considered as an alternative, as it offers twice Pure-Aloha's packet success rate. It however requires synchronization accross the nodes. In this paper, we propose a new slot structure adapted to devices with low quality clock, and a duty-cycle efficient synchronization protocol for LoRaWAN class A devices with the lowest overhead to date. We discuss the conditions of its integration into LoRaWAN. The experimental results confirm that it succeeds in tracking each device's synchronization state, identifying the exact moment they desynchronize and resynchronizing them. The proposed protocol is also proven to be more duty-cycle efficient than existing fixed-rate synchronization solutions.
\end{abstract}

\IEEEpeerreviewmaketitle

\section{Introduction}
Since its introduction almost a decade ago, the Long Range Wide Area Network (LoRaWAN) technology has met an unprecedented success among existing Low Power Wide Area Networks (LPWAN) technologies. Its low cost, low power and ability to operate over noisy channels made it reach several markets and application sectors like smart cities, smart grids and metering, agriculture monitoring, wireless sensor networks and localization \cite{2022_lorawan_smart_cities, 2018smart_grid_and_metering, 2017_agriculture_wsn_lorawan, 2018lorawan_tdoa_localization, etiabi2023spreading-localization}.
With the increasing number of nodes, LoRaWAN enters the so-called Massive IoT era. Therefore, scalability is currently one of the biggest research challenges \cite{jouhari_survey_scalable_lorawan}.

To cope with the previous challenges, propositions include the use of slotted-Aloha as an alternative to pure-Aloha which LoRaWAN currently uses. It requires all the nodes to be synchronized. LoRaWAN class B devices regularly receive beacons from the gateway that can be used to synchronize. Similarly, class C devices are listening to the channel almost all the time, hence can be synchronized using broadcast signals. The challenge resides with class A devices that do not share any synchronization signals. Several works have targeted synchronization mechanisms in wireless sensor networks over the past years  \cite{phan2022-time-sync-wsn}\cite{huan2022timestamp-sync-wsn}, but very few were dedicated to LoRaWAN, \cite{2018slotted_aloha_synchronizatio_lorawan, polonelli2019slotted-aloha-on-lorawan, energy-effi-lorawan-slotted-aloha}. Nonetheless, they didn't take into account the duty cycle limitation on LoRaWAN gateways, whilst this is a key factor that requires optimization for a wide slotted-Aloha adoption in LoRaWAN.

With that in mind, we propose in this paper a protocol that tracks the synchronization state of end-devices based on their uplink packets arrival time. This allows the network server to detect the specific moment a device desynchronizes and transmits, through the acknowledgement packet, the necessary information for its resynchronization. This minimizes the downlink overhead and results in a duty cycle gain for the gateway, without requiring any additional hardware. This paper focuses on the synchronization aspect rather than a slotted-Aloha implementation. Our contribution is twofold: 
\begin{itemize}
    \item We introduce a two-guard-intervals variant of the existing slot structure. This design allows us to define the acceptable bounds on end-devices drift, for a given length of the guard intervals.
    \item We propose a monitoring and synchronization protocol on top of 
    LoRaWAN
    that is able to identify any desynchronized devices and resynchronize them. We show that it requires only two additional bytes to encode the synchronization-related information.  To the best of our knowledge, this is the LoRaWAN synchronization protocol with the lowest overhead.
\end{itemize} 
The rest of this paper is organized as follows: Section II states the problem after a short LoRaWAN overview; Section III introduces the proposed synchronization protocol and Section IV describes how it integrates into the existing LoRaWAN stack; Section V presents and discusses the experimental results of our synchronization solution; Section VI concludes the paper and set forth our perspectives.

\section{ LoRaWAN Overview and Problem Statement}
\begin{figure}[!t]
    \centering
    \includegraphics[scale=0.25]{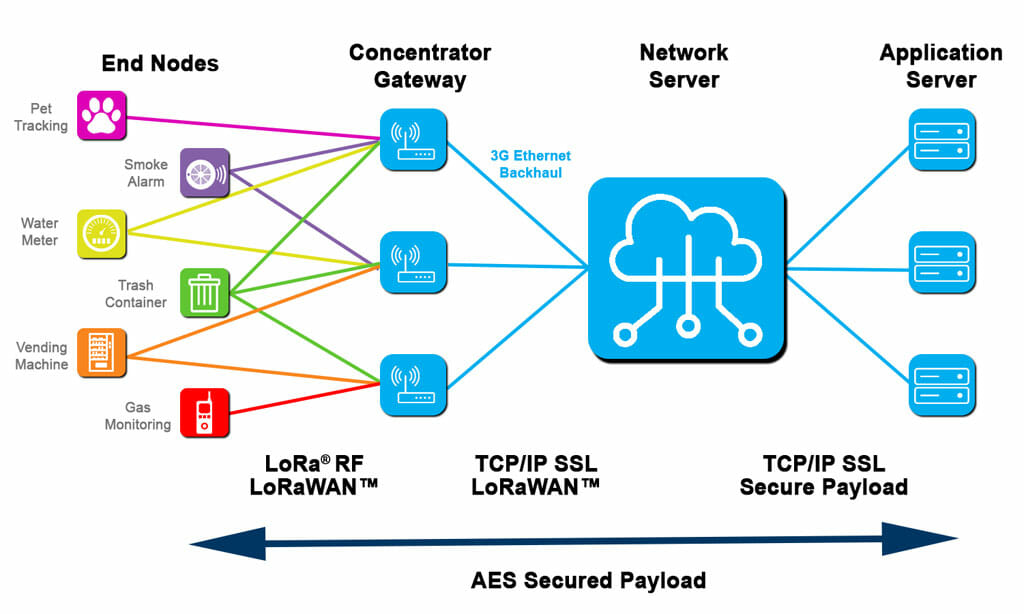}
    \caption{LoRaWAN Star of Stars topology}
    \label{fig:lorawan-topology}
    \vspace{-10pt}
\end{figure}

LoRa (Long Range) is a proprietary modulation technique developed by Semtech Corporation. It is based on the Chirp Spread Spectrum (CSS) technique, and operates in the sub-Gigahertz frequency bands. LoRaWAN, on the other hand, is a communication protocol built on top of LoRa, and designed specifically for wide-area networks with low-power consumption, low datarate IoT devices. It was developed by the LoRa Alliance \cite{lorawan_specs}. It employs a star of stars architecture, as depicted in Fig. \ref{fig:lorawan-topology}: End-devices are served by one or more gateways which, in turn, communicates with a network server. The end-devices transmit their frames to the gateways using the radio interface; the gateways forward those frames to the network server through a back-haul link; the network server is in charge of sending the payload to the intended application server. Similarly, in case of a downlink, the network server sends the frame to one gateway that forwards it to the intended end-device through the radio interface. Additionally, three classes of end-devices are supported, corresponding to different power saving policies: class B and C are respectively for deterministic and lowest downlink latency, but consume more power; class A, on the other hand, has the lowest power consumption profile, but does has a non-deterministic downlink latency since the only available downlink windows are after an uplink transmission initiated by an end-device.

Pure-Aloha is an access scheme that allows devices to transmit anytime without any form of coordination. Because of its simplicity, it is adapted to resource-constrained devices and is often adopted by low-power and low-complexity networks. It is however limited to only 18\% packet success rate when the number of nodes increases, due to collisions \cite{abramson1970aloha}. On the other hand, slotted-Aloha doubles the packet success rate, but requires the nodes to be synchronized. LoRaWAN was initially based on pure-Aloha but, with the increasing number of end-devices, slotted-Aloha is being seen as a good alternative. 

LoRaWAN targets low-cost end-devices, that can have poor clock quality, leading to significant time drift. Hence, devices should be resynchronized on regular basis to avoid slot violation. Existing synchronization solutions use the end of uplink transmission as a common reference between an end-device and the network server \cite{2018slotted_aloha_synchronizatio_lorawan} \cite{polonelli2019slotted-aloha-on-lorawan}: the network server appends the timestamp of the end of uplink into the downlink sent back to the end-device; the latter will compare this value to the one it locally stored and use the difference to correct its clock drift. This however results in a greater downlink air-time, thus consumes more duty cycle. Consequently, the challenge resides in finding a synchronization rate that keeps the end-devices drift under a certain threshold, while minimizing the duty cycle overhead.
Assessing each node's clock quality to set the synchronization rate accordingly is not practical: one network could serve thousands of nodes with different clock qualities and this operation would take a huge amount of time; plus, it will require a non-negligible software overhead to match each node with its synchronization rate.
On the other hand, choosing a single synchronization rate for all nodes seems practical, but is not optimal: nodes with better clock quality will be resynchronized more often that required, adding unnecessary overhead, while those with worse clock quality will not be resynchronized on time, leading to significant drift, hence slot violation. Moreover, a node's drift is also subject to external conditions such as voltage level and temperature. Hence, a static synchronization becomes very limited facing the previous challenges.

In what follows, we propose a synchronization protocol that dynamically adapt to each node by tracking its drift in time.

\section{Dynamic Synchronization Protocol Design}
In this section, we describe the layout of the proposed synchronization protocol, starting by detailing the underlying slot structure.

We consider a LoRaWAN gateway (GW) connected to a network server (NS) and serving $N$ class A end-devices (ED). We note $t_{ns}(t)$ the local time of NS at a moment t and $t_{ed}^{k}(t)$ that of the k-th ED, $k = \{0, 1, ..., N-1\}$. Network servers run on dedicated hardware with much more stable clock compared to end-devices. So, for the remainder of this paper, we will consider the NS clock perfect. We also suppose that NS and the EDs share the same time origin. Thus, at any moment t, the time drift $\Delta t^{(k)}$ of the k-th ED is given by Eq. \eqref{eq:time_drift}.
\begin{equation}
    \Delta t^{(k)}(t) = t_{ns}(t) - t_{ed}^{(k)}(t)
    \label{eq:time_drift}
\end{equation}

\begin{equation}
    |\Delta t^{(k)}(t)| \leq \Delta_{max}
    \label{eq:time_drift_max}
\end{equation}

Given a maximum tolerable drift $\Delta_{max}$, an ED is considered \textit{in-sync} if the condition in Eq. \eqref{eq:time_drift_max} holds, and \textit{out-sync} otherwise. If the NS can track a ED's drift accross time, it can then detect the exact moment it goes \textit{out-sync} and issue a resynchronization.

\subsection{Slot Structure Design}
\begin{figure}
    \centering
    \includegraphics[scale=0.5]{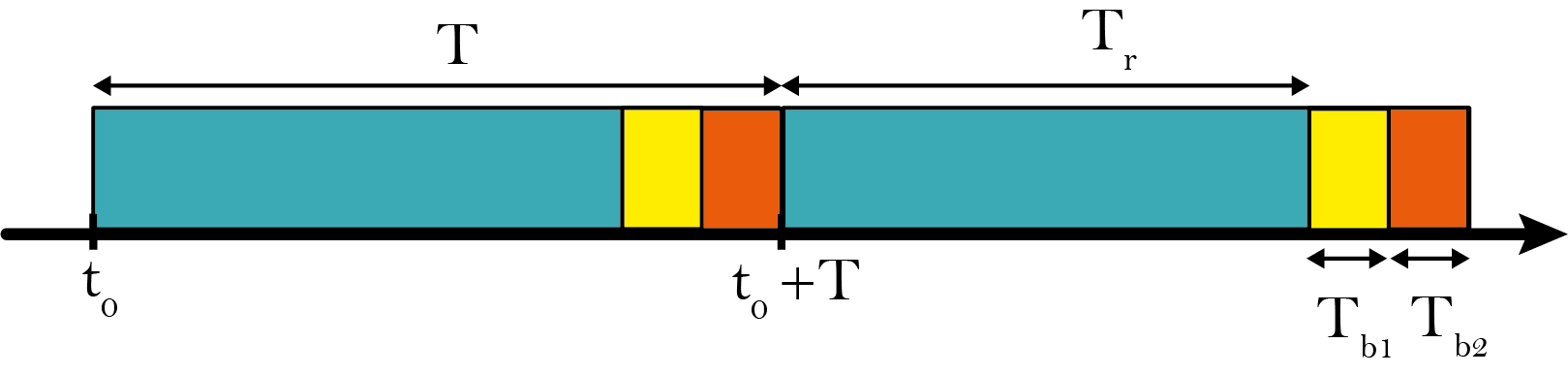}
    \caption{Slotted-Aloha: Virtual double confidence intervals}
    \label{fig:novel-slot-structure}
    \vspace{-10pt}
\end{figure}

With slotted-Aloha, time is divided into slots of predefined length and nodes can transmit only at the beginning of each slot. The classic slot consists of a transmission interval $T_r$ and a confidence interval $T_b$: the device's transmission normally starts at the beginning of the slot and lasts $T_r$; $T_b$ ensures that the transmission ends within the slot duration even in case of clock drift. As depicted by Fig. \ref{fig:novel-slot-structure}, $T_b$ can be seen as virtually made of two guard intervals: a backward guard interval $T_{b1}$ -- that prevents against drift towards the slot's start -- and a forward guard interval $T_{b2}$ -- that prevents against drift towards the slot's end. Moreover, the propagation delay is neglected: uplink air-times range from hundreds of ms while the propagation delay equals $50 \mu s$, assuming a line of sight and a typical LoRa coverage of 15 km; it can then be neglected even in case of multipath propagation. The slot is therefore made of the uplink air-time $T_{tx}$ and the confidence interval $T_b$.

Devices -- that are microcontroller-based -- keep time by counting periodic pulses generated from the system clock, which frequency $f_0$ is known. The clock however has frequency precision, depending on its quality \cite{tjoa2004clock-drift}. The instantaneous frequency is given by Eq. \eqref{eq:frequency_drift}. $\Delta f$ is a real number, so the frequency can either increase or decrease. 

\begin{equation}
    \begin{tabular}{c}
         $f(t) = f_0 + \Delta f(t)$, \\
    $\Delta f(t)$ \mbox{being the instantaneous frequency drift}.
    \end{tabular}
    \label{eq:frequency_drift}
\end{equation}

\begin{figure}[!t]
    \centering
    \includegraphics[scale=0.5]{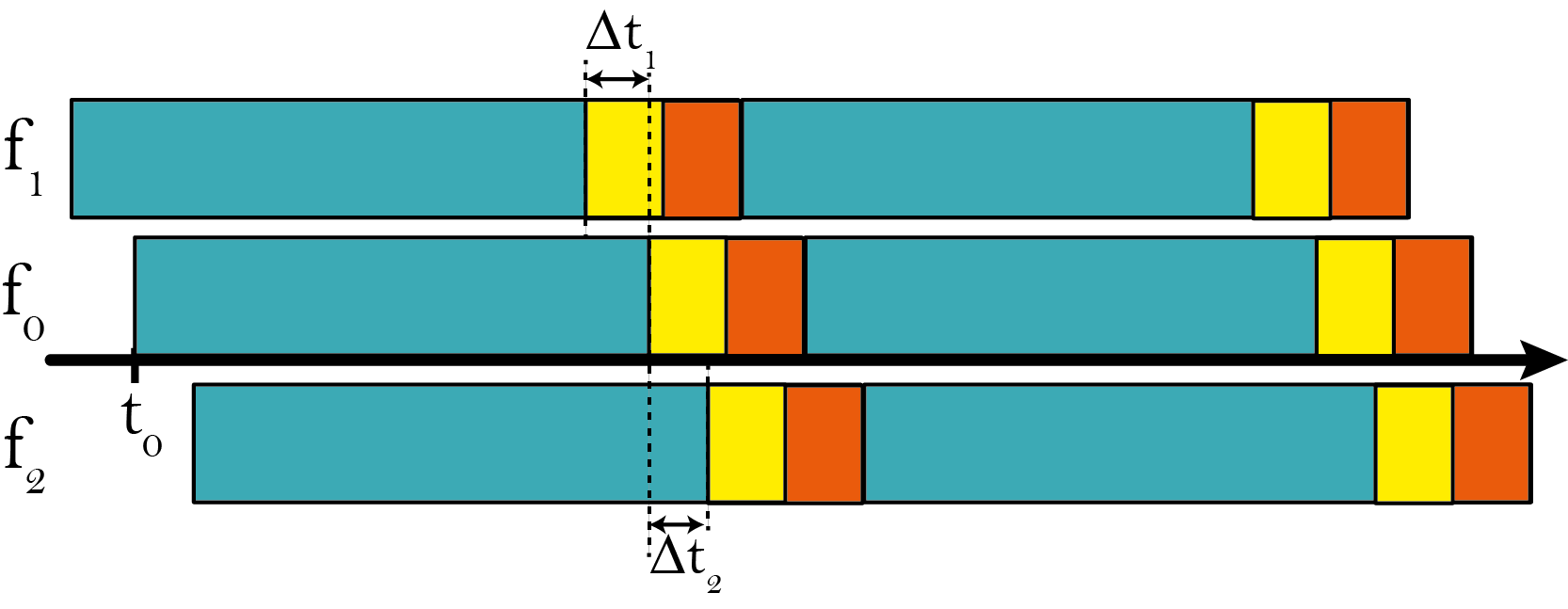}
    \vspace{-10pt}
    \caption{Apparent drift effect on slot position
    }
    \label{fig:slotted-aloha-3-slotlines}
    \vspace{-10pt}
\end{figure}

Moreover, relatively to a device with an ideal clock running at $f_0$, another device running at $f_1 > f_0$ will be ahead in time while another one running at $f_2 < f_0$ will be late, as illustrated by Fig. \ref{fig:slotted-aloha-3-slotlines}. The maximum tolerated backward and forward drifts being respectively $T_{b1}$ and $T_{b2}$, the frame \textit{in-sync} condition can be derived through Eq. \eqref{eq:in-sync-condition}. Moreover, Eq. \eqref{eq:in-sync-condition-ttx} gives the \textit{in-sync} condition if the end of uplink transmission is used as a reference. Therefore, the NS can compute the k-th ED's drift upon a frame reception, where $T_{Tx-ns}$ is the end of uplink transmission expected by the NS and $T_{Tx-ed}$ the moment the uplink transmission effectively ends. If an \textit{out-sync} is detected, the device is resynchronized using the algorithm described in the following subsection.

\begin{equation}
    \begin{tabular}{cl}
         & $ -T_{b2} < \Delta t^{(k)}(t) < T_{b1} $ \\
       $ \implies $  & $-T_{b2} < t_{ns}(t) - t_{ed}^{(k)}(t) < T_{b1} $ \\
       $ \implies $ & $ t_{ns}(t) - T_{b1} <  t_{ed}^{(k)}(t) < t_{ns}(t) + T_{b2} $
    \end{tabular}
    \label{eq:in-sync-condition}
\end{equation}
    
\begin{equation}
     T_{Tx-ns} - T_{b1} <  T_{Tx-ed}^{(k)} < T_{Tx-ns} + T_{b2}
     \label{eq:in-sync-condition-ttx}
\end{equation}

The uplink air-time $T_{tx}$ depends on physical layer parameters and can be calculated from Eq. \eqref{equ:time-on-air-1}, \cite{sx1272-73_datasheets}. Table \ref{tab:lora-params-meaning} describes the symbols involved in the equations.

\begin{equation}
    T_{packet} = T_{preamble} + T_{payload}
    \label{equ:time-on-air-1}
\end{equation}

\begin{equation}
    T_{preamble} = (n_{preamble} + 4.25)T_s
\end{equation}

\begin{equation}
    T_s = \frac{2^{SF}}{BW}
\end{equation}

\begin{equation}
    T_{payload} = n_{payload}\times T_s
\end{equation}

\begin{equation}
    \begin{tabular}{l c c}
         $n_{bits}$ & $=$ & $ ceil\left(\frac{8PL - 4SF + 28 + 16CRC - 20IH}{4(SF-2DE)}\right)$ \\
         $n_{payload}$ & $=$ & $ 8 +  max \left(n_{bits}(CR+4), 0\right)$
    \end{tabular}
\label{equ:time-on-air-5}
\end{equation}

\begin{table}[!t]
    \caption{Air-time equations symbols description}
    \label{tab:lora-params-meaning}
    \centering
    \begin{tabular}{|l|m{20em}|}
    \hline
        Symbols  &       Parameters     \\
    \hline
        Tpacket   &  Total packet air-time \\
        Tpreamble   & Preamble duration         \\
        Tpayload &    Payload duration      \\
        Ts          &   Symbol period   \\
        npayload    &   Number of payload symbols   \\
        PL                          &   Payload size (in bytes)     \\
        SF                          &   Spreading Factor            \\
        IH                          &   Implicit Header flag        \\
        DE                          &   Low data rate optimization flag \\
        CRC                         &   Cyclic Redundancy Check flag \\
        CR                          &   Coding Rate                  \\
    \hline
    \end{tabular}
    \vspace{-10pt}
\end{table}

The guard intervals $T_{b1}$ and $T_{b2}$ represent unconsumed time slots, so they should be calculated as an acceptable ratio of $T_{tx}$, based on the application.

\subsection{Synchronization and Monitoring Protocol}

\begin{algorithm}
    \centering
     \caption{Network Server} \label{alg:sync-ns}
    \begin{algorithmic}[1]
        \State{receive()}
        \State{arrival $\gets$ current\_time() - ref}
        \State{arrival $\gets$ arrival mod T\_slot}
        \If{out\_sync()}
            \State{remain $\gets$ T\_slot - arrival}
            \State{add\_remain\_to\_ack()}
        \EndIf
        \State{wait\_for\_rx\_window()}
        \State{transmit\_ack()}
    \end{algorithmic}
\end{algorithm}
\begin{algorithm}
    \centering
    \caption{end-device}\label{alg:sync-end-device}
    \begin{algorithmic}[1]
        \If{!is\_first\_tx}
            \State{is\_first\_tx} $\gets$ 1
            \State{slot\_start $\gets$ current\_time()}
        \Else
            \State{wait\_for\_next\_slot()}
        \EndIf
        \State{transmit()}
        \State{beg $\gets$ current\_time()}
        \State{wait\_for\_rx\_window()}
        \State{get\_ack()}
        \State{end $\gets$ current\_time()}
        \If{remain\_in\_ack()}
            \State{remain $\gets$ get\_remain\_from\_ack()}
            \State{elapsed $\gets$ end - beg}
            \State{t $\gets$ remain - elapsed}
            \If{$t < 0$}
                \State{t $\gets$ t mod T\_slot}
            \EndIf
            \State{slot\_start $\gets$ current\_time() + t}
        \EndIf
        
    \end{algorithmic}
\end{algorithm}

The core idea of our protocol lies in a unique time reference, \textbf{ref}, stored at the network server and is used to synchronize all the end-devices and later monitor their synchronization from the incoming packets arrival time. \textbf{ref} is set only once with the current time when the network server boots up, and it is considered as the beginning of the very first slot. From there it is easy to get the future slots start through Eq. (\ref{equ:slot_start}). 

\begin{equation}
    \begin{tabular}{c r}
        $ slot\_start[n] = ref + n \times T\_slot$, &   $n \in \mathbf{N^*}$ \\
    \end{tabular}
    \label{equ:slot_start}
\end{equation}

At any time, Eq. (\ref{equ:position_in_slot}) gives the relative position in the ongoing slot -- that is, how much time elapsed since the beginning of this slot. Similarly, Eq. (\ref{equ:remaining_to_slot}) gives the remaining time before the next slot.
\begin{equation}
     position = (time - ref) \mbox{ mod } T\_slot 
    \label{equ:position_in_slot}
\end{equation}

\begin{equation}
    remaining\_time =  T\_slot - ((time - ref) \mbox{ mod } T\_slot)
    \label{equ:remaining_to_slot}
\end{equation}

The protocol is divided into two algorithms: one for the NS and the other one for EDs. For an ED, the synchronization consists in simply saving the beginning of a slot into a variable at the right time, so that it can serve to identify future slots start using Eq. (\ref{equ:slot_start}). To keep the algorithms short, we substituted some portions with self-explanatory function names such as \textbf{current\_time()}, \textbf{wait\_for\_next\_slot()}, or \textbf{wait()}. Those are just steps of the algorithms that can be simple sets of instructions instead of functions. Below, we explain the flow of both algorithms.

\subsubsection{Algorithm 1: Network Server}
Lines 1-3: The server waits for incoming packets. Once a packet is received, the server immediately uses Eq. (\ref{equ:position_in_slot}) to calculate its position inside the ongoing slot. \\
Lines 4-7: The server checks whether the packet is \textit{out-sync}; in which case, it gets the time remaining before the next slot from Eq. (\ref{equ:remaining_to_slot}), and piggybacks it to the ACK packet. \\
Lines 8-9: Once the ACK packet is ready, the server waits for the next receive window and sends it.

\subsubsection{Algorithm 2: end-device}
Lines 1-6: An ED that transmits for the first time does not have any synchronization reference on its side. So, it takes the transmission moment as the reference. If that first frame arrives \textit{in-sync}, this reference will be kept, otherwise it will be corrected. The corrected reference will be used to determine slot start for future transmissions (line 5). The flag \textbf{is\_first\_tx}, by default initialized to \textbf{false}, is used to identify the first transmission. \\
Lines 8-11: The end of the transmission is timestamped into \textbf{beg}. This instant coincides with the one when the NS takes the packet arrival time. The moment the ACK packet is received from the NS is also timestamped in \textbf{end}. \\
Lines 12-20: If the received ACK packet contains remaining-time bytes, it means the last transmission was received \textit{out-sync}. The remaining time is extracted and used to resynchronize the device. The elapsed time between the moment the network server computed the remaining-time and the moment the device received the ACK packet -- equal to \textbf{end - beg} -- is substracted from remaining-time. Then, The ED resynchronizes by updating its slot start reference with remaining-time added to its current local time. It can happen that the time elapsed is greater than the remaining time. This scenario is handled in lines 16-18 by the mean of a modulo operation. Additionally, the presence of remaining-time in the ACK packet can be checked by different means: for instance, the implementation used for the experimental results section below assumed a downlink with no MAC commands; so, checking the packet size was enough in that case. 

\section{Protocol Integration into LoRaWAN}
With the protocol already established, we now discuss how it integrates into the LoRaWAN stack. LoRaWAN frame structure includes a Frame Options field for sharing MAC commands between end-devices and the network server. We use this field to send the synchronization-related bytes, the \textbf{remaining-time} for instance. Its value is between 0 -- for a frame received exactly at the end of the slot -- and $T_{slot}$ -- for a frame received at the exact beginning of a slot. Table \ref{tab:longest-time-on-air-params} specifies the radio parameters that gives the longest packet air-time: LoRa spreading factors ranges from 5 to 12, the greater the spreading factor the more the signal is spread in time; LoRa defines coding rates 1, 2, 3 and 4 corresponding resp. to effective rates $\frac{4}{5}$, $\frac{4}{6}$, $\frac{4}{7}$ and $\frac{4}{8}$; LoRaWAN uses only 125 kHz, 250 kHz and 500 kHz bandwidth, the smallest bandwidth giving the greatest symbol duration thus the longest air-time; LoRa allows a payload limit of 255 bytes. The resulting maximum uplink air-time equals 11936 ms, requiring $\log_2(11936)$ = 14 bits for encoding. 2 bytes are therefore enough to encode the slot length $T\_slot$ and leave a margin of 53599 ms for $T_{b1} + T_{b2}$. Those 2 bytes are the overhead induced by our protocol, against the 8 bytes in \cite{2018slotted_aloha_synchronizatio_lorawan}.

To enable coexistence of devices using this protocol for slotted-Aloha with the preexisting ones in LoRaWAN, we propose to dedicate special FPort (Frame Port) values to this protocol in LoRaWAN specifications. This is an efficient method, as it will allow the NS to distinguish the ED requiring synchronization from those that don't, and this without any additional overhead on the transmitted payload. In the performance evaluation section, we set the FPort to 198; it is not used for any services in the current specification, except for proprietary protocols.

\begin{table}[!t]
    \caption{Longest air-time LoRa Radio Parameters.}
    \label{tab:longest-time-on-air-params}
    \centering
    \begin{tabular}{|l|c|}
    \hline
        Parameter  & Value \\
     \hline
        Coding Rate &   4   \\
        Spreading Factor    &   12  \\
        Bandwidth (kHz)   &   125 \\
        Payload  & 255   \\
    \hline
    \end{tabular}
\end{table}

\begin{figure}[!t]
    \centering
    \includegraphics[scale=0.105]{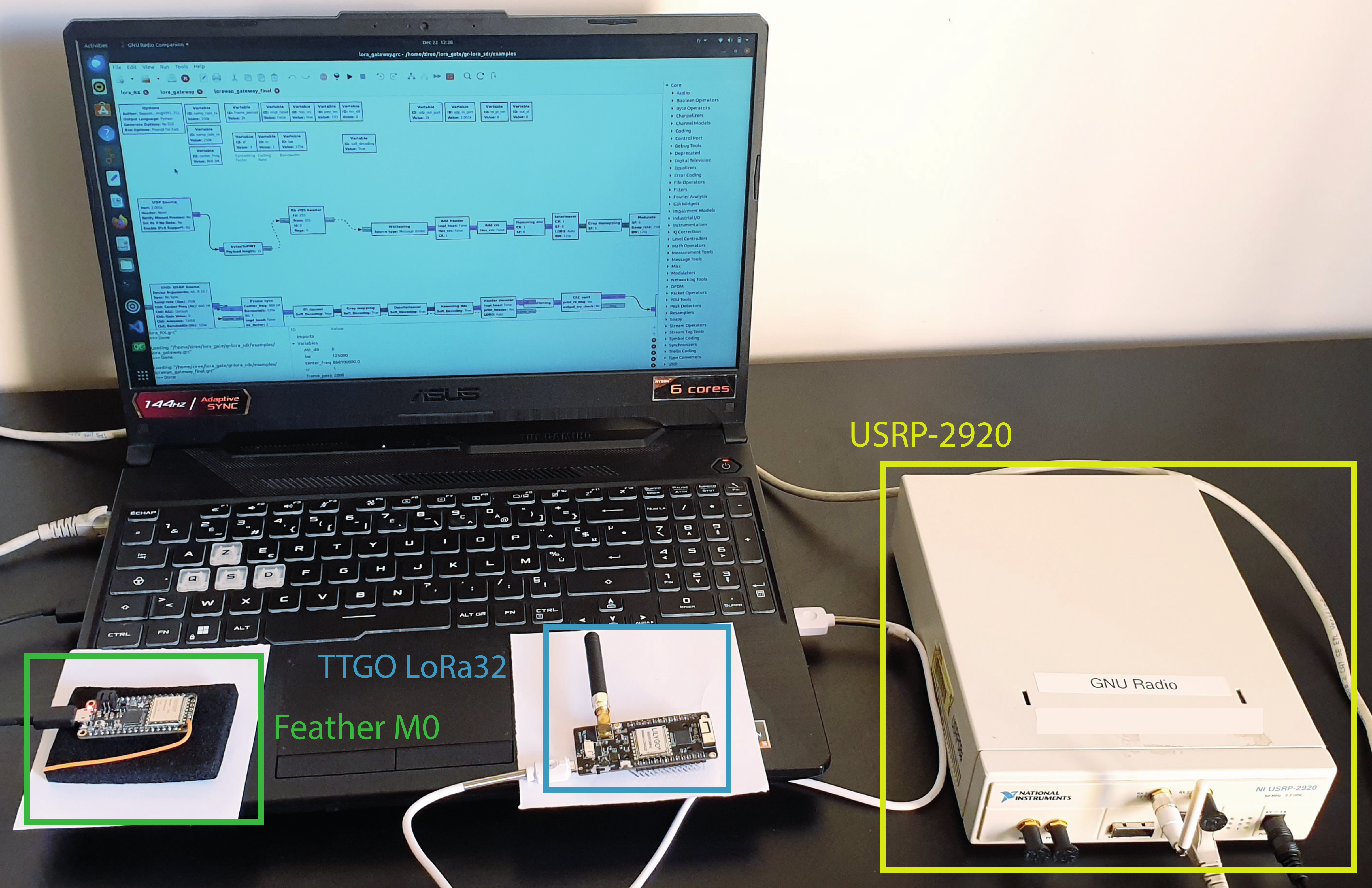}
    \caption{Test bench equipment: A National Instruments USRP-2920 SDR as Gateway; An Adafruit Feather M0 and a TTGO Esp32 LoRa are used as end-devices; The network server was running on a 6-cores 12-threads 4.0 GHz laptop.}
    \label{fig:test-bench}
    \vspace{-10pt}
\end{figure}

\section{Experimental Results}
\begin{figure}[!t]
    \centering
    \includegraphics[scale=0.2]{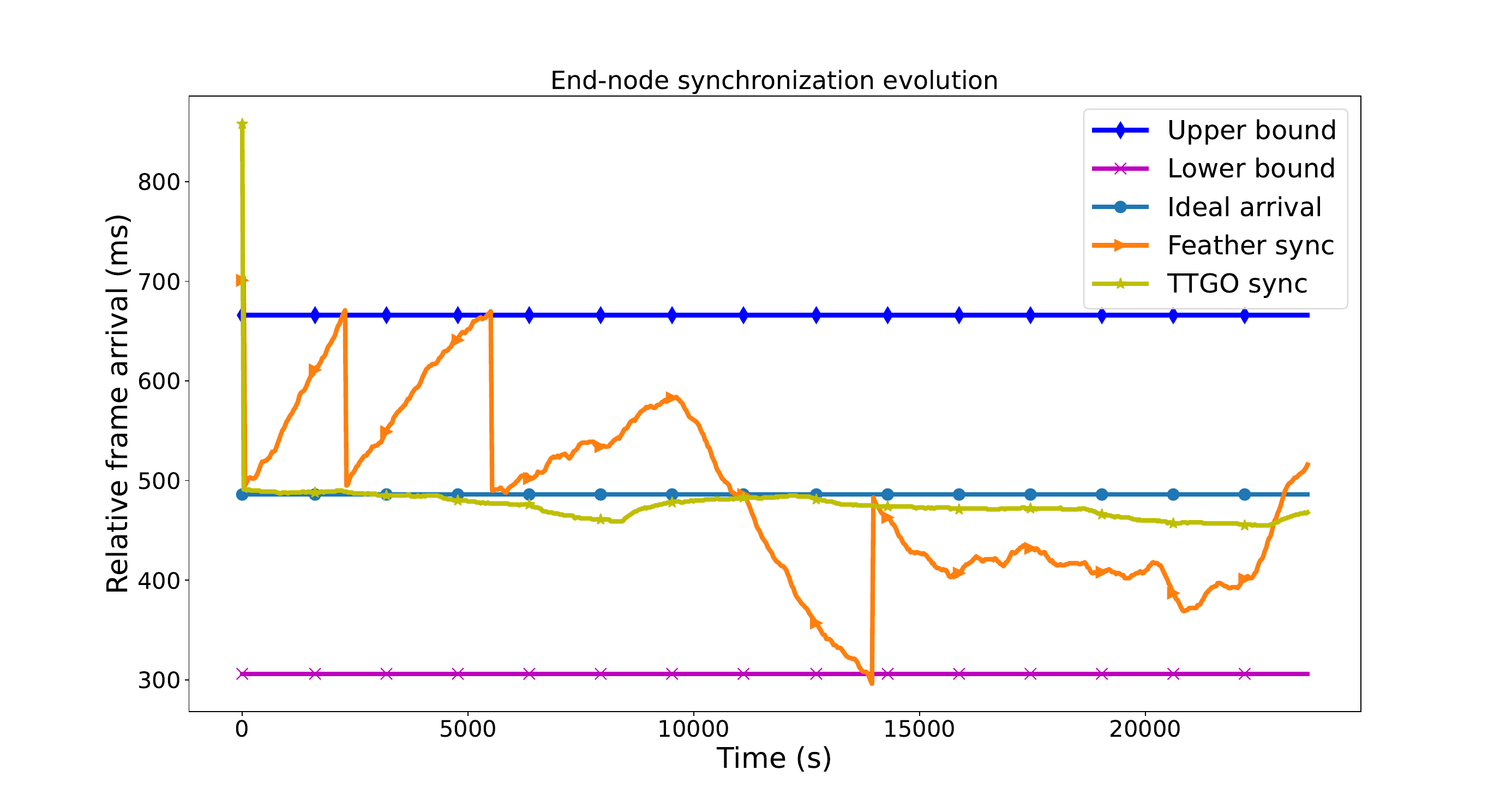}
    \vspace{-20pt}
    \caption{Protocol performance: devices synchronization and monitoring}
    \label{fig:sync_evolution}
    \vspace{-10pt}
\end{figure}

\begin{figure}
    \centering
    \subfloat[]{\includegraphics[scale=0.35]{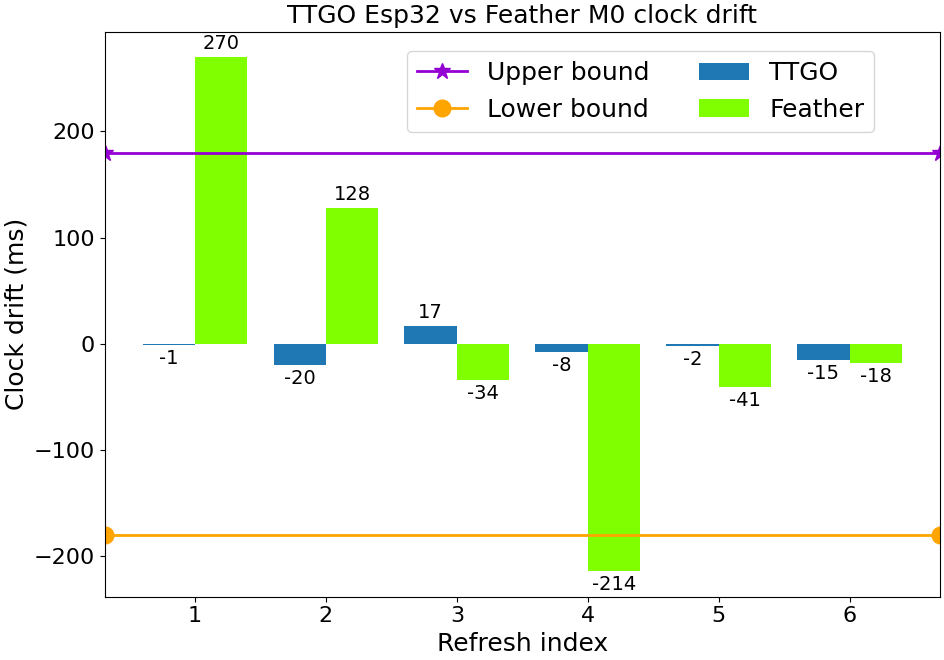} \label{fig:fixed-rate-1h}}
    \hfill
    \subfloat[]{\includegraphics[scale=0.35]{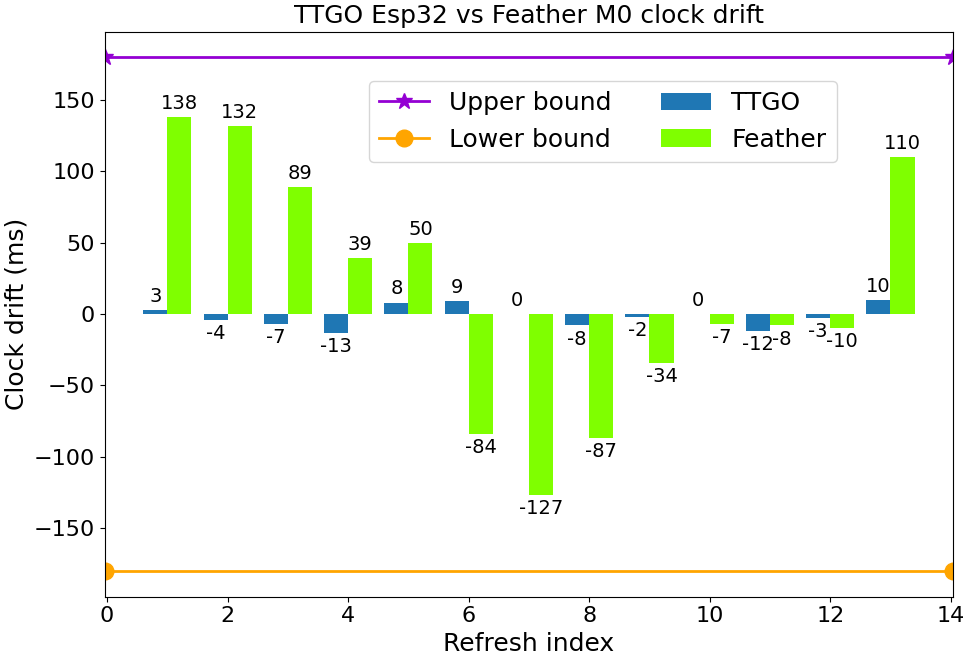} \label{fig:fixed-rate-30min}}
    \caption{Fixed-rate synchronization algorithm performance, 6.5 hours experiment: (a) 1 hour round duration; (b) 30 mins round duration}
    \label{fig:fixed-rate-sync}
    \vspace{-10pt}
\end{figure}

In this section, we set up a test bench to evaluate the performances of our protocol. We set up a custom gateway by providing few additional components to the Open Source GNU Radio implementation of LoRa physical layer in \cite{tapparel_lora_gnuradio}. We also developed a LoRaWAN network server with the necessary mechanisms for this protocol. All the related codes are made available on Github \cite{code-github-link}. Fig. \ref{fig:test-bench} shows our experimental test bench: a USRP-2920 serves as a gateway, and an Adafruit Feather M0 -- that has a very unstable clock -- and a TTGO Esp32 LoRa -- with a much more stable clock -- boards were used as end-devices. The network server was running locally, on a 6-cores 4.0 GHz laptop.

The performance of our protocol will be compared to that of the algorithm used in \cite{2018slotted_aloha_synchronizatio_lorawan}, where upchirp-modulated symbols were used for both the uplink and downlink; as a consequence, the slot was made of the uplink air-time $T_{tx}$, the downlink air-time $T_{rx}$, the 1s gap, and the guard interval $T_{b1}$ and $T_{b2}$. So, we did the same here for a fair comparison.

We first run an experiment using the proposed synchronization protocol. The end-devices transmit a packet every 30s. The parameters in Table \ref{tab:experiment_lora_params} together with Eqs. (\ref{equ:time-on-air-1}-\ref{equ:time-on-air-5}) gives the air-time in Table \ref{tab:experiment_time-on-air}. Total slot length refers to the slot duration and accounts for both the uplink and downlink air-time, the 1000 ms delay, and the slot guards $T_{b1}$ and $T_{b2}$ of 180 ms each. We run the setup for 6.5 hours, logged the relative position of each frame within the ongoing slot, and plotted the evolution of the devices synchronization in Fig. \ref{fig:sync_evolution}. The upper and lower bounds mark the limit of the synchronization interval, and the ideal arrival line marks the position of frames received from a perfectly synchronized node. The first frames of both devices were received out-of-sync; that was expected since they have a chance of only $\frac{T_{b1}+T_{b2}}{T_{slot}} \approx$ 20\% to be received \textit{in-sync}. They got synchronized immediately and their next transmission matches the ideal arrival line. The evolution of the synchronization graphs shows the difference in the devices' clocks stability and the success of the protocol in resynchronizing them: The feather desynchronized three more times -- twice during the first two hours -- and get corrected right after, while the TTGO never did because its drifts are very small compared to the guard intervals value; hence it didn't need any resynchronization. In total, both devices were synchronized 5 times during the 6.5 hours.

We run another two other experiments, this time using the fixed-rate synchronization algorithm described in \cite{2018slotted_aloha_synchronizatio_lorawan}: the EDs transmit a packet every 30s; after each round, the NS measures their clock drift, logs it and re-synchronizes them; each round last 1 hour for the first experiment and 30 minutes for the second one. The resulting drifts are plotted as comparative bar diagrams between both devices and for both experiments. For the 1 hour round duration in Fig. \ref{fig:fixed-rate-1h}, the fixed-rate algorithm encounters two slot violations by the Feather and none by TTGO, and both were resynchronized in total 12 times during the whole experiment; this is 2.4 times more overheads than our adaptive algorithm that does even prevent slot violation . And, for the 30 mins round duration from Fig. \ref{fig:fixed-rate-30min}, there has been no slot violation for any of the devices, but they were however resynchronized 26 times; this is five times more overheads than our algorithm and for the same result.

In summary, it clearly appears that the fixed-rate synchronization algorithm is not able to prevent slot violations while keeping a low synchronization overhead at the same time. Our algorithm, on the other hand is able to accomplish both by adapting to each device separately, hence succeeds in being more duty-cyle efficient. 

\begin{table}[!t]
    \centering
    \caption{Experiment LoRa radio parameters}
    \label{tab:experiment_lora_params}
    \begin{tabular}{|l|c|}
        \hline
         Parameter    & Value \\
        \hline
         Frequency (MHz)    & 868 \\
         Bandwidth (kHz)    & 125\\
         Coding Rate  &  1   \\
         Preamble size & 8  \\
         Uplink Payload size   &   193 \\
         Uplink Spreading Factor & 7   \\
         Downlink Payload size   &   19  \\
         Downlink Spreading Factor   &   8   \\
        \hline
    \end{tabular}    
    \vspace{-10pt}
\end{table}

\begin{table}[!t]
    \centering
    \caption{Experiment Packets air-time}
    \label{tab:experiment_time-on-air}
    \begin{tabular}{|l|c|}
        \hline
           Parameter  & Value (ms)  \\
        \hline
            Uplink air-time & 306 \\
            Downlink air-time & 91 \\
            Total slot length    & 1757 \\
        \hline
    \end{tabular}
    \vspace{-10pt}
\end{table}

\section{Conclusion}
In this paper, we proposed and tested a synchronization protocol for LoRaWAN class A devices, designed to optimize gateways duty cycle consumption by reducing the downlink air-time overhead, along with a novel slot structure. The experimental results confirmed that our protocol was able to track each device synchronization state separately, and resynchronize it only when it detects a desynchronization. Furthermore, its low overhead (only two bytes), duty-cycle efficiency and seamless integration with existing schemes position it as a powerhouse for slotted-Aloha adoption in LoRaWAN deployments within the current massive IoT context.

Our solution could be improved by reducing downlink ACK packets, since it currently requires a downlink acknowledgement for each uplink so that devices can identify collisions and retransmit if necessary. Furthermore, future works will investigate the impact of the increasing number of nodes on the synchronization accuracy.
\section*{Acknowledgment}
This work was sponsored by the Junior Faculty Development program under the UM6P-EPFL Excellence in Africa
Initiative.
\vspace{-0.05in}


\bibliographystyle{IEEEtran}
\bibliography{IEEEabrv,references}

\end{document}